\RequirePackage{fix-cm}

\documentclass[smallextended]{svjour3}       
\smartqed  
\usepackage{amssymb}
\usepackage{graphicx}
\usepackage[inner=2.5cm,outer=2cm]{geometry}
\usepackage{algorithm}
\usepackage{algpseudocode}
\usepackage{algpascal}
\begin{document}

\title{Approximation Algorithms for the Load Balanced Capacitated Vehicle Routing Problem}



\author{Haniyeh Fallah  \and
        Farzad Didehvar$^*$  \and
        Farhad Rahmati  
}


\institute{$^*$Corresponding author: Farzad Didehvar$^1$ \at
                   \email{didehvar@aut.ac.ir}       
           \and
        Haniyeh Fallah$^1$  \at
              \email{hfallah@aut.ac.ir} 
             \and
            Farhad Rahmati$^1$ \at
          \email{frahmati@aut.ac.ir} 
          \and
          1.  Department of Mathematics and Computer Science, Amirkabir University of Technology, Tehran, P. O. Box: 15875-4413, Iran} 

\date{Received: date / Accepted: date}

\maketitle
\begin{abstract}
\noindent We study the load balanced capacitated vehicle routing problem (LBCVRP): the problem is to design a collection of tours for a fixed fleet of vehicles with capacity $Q$ to distribute a supply from a single depot between a number of predefined clients, in a way that the total traveling cost is a minimum, and the vehicle loads are balanced. The unbalanced loads cause the decrease of distribution quality especially in business environments and flexibility in the logistics activities. The problem being NP-hard, we propose two approximation algorithms.\\
\noindent When the demands are equal, we present a $((1-\frac{1}{Q})\rho +\frac{3}{2})-$approximation algorithm that finds balanced loads. Here, $\rho$ is the approximation ratio for the known metric traveling salesman problem (TSP). This result leads to a $2.5-\frac{1}{Q}$ approximation ratio for the tree metrics since an optimal solution can be found for the TSP on a tree. We present an improved $2-$approximation algorithm. \\
 \noindent When the demands are unequal, we focus on obtaining approximate solutions since finding balanced loads is NP-complete. We propose an algorithm that provides a $4-$approximation for the balance of the loads. \\
\noindent We assume a second approach to get around the difficulties of the feasibility. In this approach, we redefine and convert the problem into a multi-objective problem. The algorithm we propose has a 4 factor of approximation. 

\keywords{Capacitated Vehicle Routing Problem\and Load Balanced Capacitated Vehicle Routing Problem \and Approximation Algorithms\and Fairness}
\end{abstract}

\section{Introduction}
\label{intro}
\subsection{Problem definition}
\indent The capacitated vehicle routing problem (CVRP) is a well-known and fundamental problem in the domain of operations research. It has many applications that are mostly related to transportation, logistics industry, communications, manufacturing, military and relief systems, etc. The CVRP is a graph problem that can be defined as follows: the quantity $d_v$ should be delivered to the client $v (v=1,...,n)$ from a single depot using a vehicles fleet with capacity $Q$. A CVRP solution is a set of tours: each tour starts and ends at the depot after visiting a set of clients in a way that the sum of the clients' demands on a tour (the tour load) does not violate the capacity constraint. Each client must be visited exactly once. The objective is to minimize the total traveling cost. A special version is when each client demand is unit (i.e. $d_v=1$ for a client $v$). In this case, the capacity restriction is given as a limit for the maximal number of clients in a vehicle tour. A unit demand CVRP is also known as the $equal \,\, demand \,\,capacitated \,\,vehicle \,\,routing \,\,problem$ (ECVRP). When the clients are allowed to be visited by more than one vehicle, we have the split delivery vehicle routing problem \cite{Dror}. This problem can be reduced to a unit demand CVRP by replacing a client $v$ with demand $d_v$ by $d_v$ clients with unit demands and zero inter-distances. For an overview of the rich literature on the CVRP, we refer the reader to \cite{Altinkemer1990,Altinkemer1987,Baldacci,Bompadre,Chandran,Frederickson,Haimovich1988,Haimovich1985,Laporte,Li,Marinakis}. \\
\indent  A variety of the CVRP considers additional optimality objectives such as the balance of the routes (see \cite{Zhou}), the balance of the vehicle loads (see \cite{Bowerman}) or the minimum number of the vehicles (see \cite{Bowerman}). Chen et al. \cite{Chen2008} studied the CVRP with load balancing and time windows. Tsouros et al. \cite{Tsouros} considered the CVRP with load and route balance. Lee and Ueng \cite{Lee} studied the CVRP with load balancing where an attempt is made to balance the workload of the drivers. Bowerman et al. \cite{Bowerman} studied the CVRP with load and route balancing. In this paper, we consider the CVRP with load balancing. We denote the unequal demand version by LBCVRP and identify the equal/unit demand version by LBECVRP.\\
\indent Load balancing in a CVRP has practical applications in situations where it is required to distribute services fairly e.g. the collection of school children on a school bus \cite{Bowerman}. In their paper, Bowerman et al. \cite{Bowerman} investigated the design of school bus routes in an urban setting in Canada. Alongside the goal of minimizing the route numbers, they also followed further goal of adjusting the students number served in each route. The authors proposed a multi-objective optimization approach that minimizes the number of routes and at the same time balances the number of students. According to their assertions, the load balancing reduces the likelihood that the routes will be overfilled with an additional load, when fresh students enter into a school attendance zone, or if school attendance zones are re-expressed and redefined when adjusting pre-existing vehicle routes. \\
\indent In this article, we provide approximation algorithms for the LBECVRP and the LBCVRP since they are NP-hard problems. As per our knowledge, these problems have not yet been addressed by the approach of approximation algorithms. We use the approach of Vazirani \cite{Vazirani} to define an approximation algorithm and approximation guarantee. An approximation algorithm obtains a feasible solution in polynomial time and guarantees the solution quality.   \\
\indent Here, we restrict ourselves to tree metrics and we refer to it as the tree load balanced capacitated vehicle routing problem (TLBCVRP). In this case, each vehicle tour could be characterized by a tree (rooted at the depot) and a vehicle may pass through a node without serving it. Naturally, tree networks emerge in situations where access construction is more expensive than routing costs e.g. in pit mine railways and river networks (see \cite{Basnet,Labbe}). Further, CVRP on trees has applications in flexible manufacturing environments \cite{Berger}. In addition to the practical applications, the TLBCVRP is compelling theoretically. We are interested to know whether this problem can be solved more efficiently than the general LBCVRP. Here, we provide an improved approximation algorithm when the demands are equal. Through this article, we denote the equal demand version by TLBECVRP.
\subsection{Our results}
\indent  We give a load balancing algorithm that finds the balanced loads, in the equal demand case (LBECVRP). Given an instance of the LBECVRP and its balanced loads, we present the first $((1-\frac{1}{Q})\rho +\frac{3}{2})-$approximation algorithm, where $\rho$ is the approximation ratio for the known metric traveling salesman problem (TSP). Our algorithm is based on a method called route-first cluster-second method  (\cite{Beasley}) for the vehicle routing problems. This result leads to a $(2.5-\frac{1}{Q})-$approximation algorithm for the TLBECVRP, since TSP can be solved optimally on trees. We present an improved $2-$approximation algorithm. This algorithm proceeds in a sequence of rounds and partitions the tree into subtrees whose loads are balanced. Since the balanced loads differ in at most once, we reform the tree and prepare some strategies to partition it (see section 3). \\
\indent In the unequal demand case, finding balanced loads is NP-complete (see section 2.3), so we provide an approximate solution. We show there is a $(4.5-\frac{3}{Q},4)$ bi-criteria approximation algorithm that finds a solution whose cost is $4.5-\frac{3}{Q}$ times the optimal and the balance of the loads is 4-approximation of the optimal balance. When the metric space is a tree, we show there is a $(3,4)$ bi-criteria approximation algorithm. In a second approach, we redefine and convert the problem into a multi-objective problem and present a $4-$approximation algorithm.
\subsection{Related works}
\indent  The LBCVRP belongs to the class of combinatorial problems, known as balanced optimization problems. Martello et al. \cite{Martello} characterized a general class of these problems, for the first time. Since then, many variations and generalizations of them were studied (see e.g. \cite{Ahuja,Aydin,Cappanera}). Larusic and Punnen \cite{Larusic} studied the balanced traveling salesman problem which has applications in modeling optimization problems where it is important to equitably distribute resources. Some other assignment is allocated to Bassetto and Mason \cite{Bassetto} where a 2-period balanced traveling salesman problem is regarded in Euclidean graphs. A variety of the balanced vehicle routing problems have been studied by authors (see e.g. \cite{Borgulya,Chen2009,Chen2008,Lee,Matl,Tsouros,Yousefikhoshbakht}). \\ 
\indent The first approximation analysis of the CVRP is done by Haimovich and Rinnooy Kan \cite{Haimovich1985}. They presented a lower bound on the optimal cost of the ECVRP and derived approximation results based on this bound. Also, they provided the first PTAS for the ECVRP on Euclidean metrics. Further research articles (e.g., \cite{Altinkemer1990,Altinkemer1987,Li}) depend on this bound to improve or generalize approximation results for the CVRP (equal or unequal demands). Altinkemer and Gavish \cite{Altinkemer1987} have proposed a $(2+(1-\frac{2}{Q})\rho)-$approximation algorithm for the CVRP which improves the result of Haimovich and Rinnooy Kan \cite{Haimovich1985}. Here, $\rho$ is an approximation factor for the known metric Traveling Salesman Problem (TSP). The best result for $\rho$ is $\frac{3}{2}$ in general metrics \cite{Christofides}, and $(1+\epsilon)$ for Euclidean metrics \cite{Mitchell}. In 1990, Altinkemer and Gavish \cite{Altinkemer1990} presented a $(1+(1-\frac{1}{Q})\rho)-$approximation algorithm for the ECVRP. Recently, Bompadre et al. \cite{Bompadre} prepared lower bounds for the CVRP (equal and unequal demands) which improves the bound of Haimovich and Rinnooy Kan \cite{Haimovich1985}. For the CVRP with multiple depots, Li and Simchi-Levi \cite{Li} have presented a $5-$approximation algorithm. A version of the CVRP called $Q-$delivery is considered by Charikar et al. \cite{Charikar}. They derive a $5-$approximation, for this problem. In the latter, a single vehicle with capacity $Q$ transports $n$ items located at arbitrary locations to given demand points. The CVRP on trees (TCVRP) is introduced by Labbe et al. \cite{Labbe}. They proved NP-hardness of this problem and presented a $2-$approximation algorithm. Another $2-$approximation algorithm is given by Chandran and Raghavan \cite{Chandran}. Hamaguchi et al. \cite{Hamaguchi} studied the split delivery vehicle routing problem on trees and provided a $1.5-$approximation algorithm. Also, they showed that it is an NP-hard problem. Asano et al. \cite{Asano2001} improved this ratio to $1.35078$. Recently, Becker \cite{Becker} presented a tight $\frac{4}{3}-$approximation for this problem. \\

\indent $\mathbf{Paper\,\,outline}.$ This paper is organized as follows. In section 2, we give a precise formulation of the problem. In section 3, we address the LBECVRP in general and on trees. We give a load balancing algorithm that finds the balanced loads. Then, we present two approximation algorithms. We study the general case for the unequal demands (LBCVRP) in general metrics and on trees, in section 4. We also discuss the multi-objective version of the LBCVRP, in this section. Finally, we summarize the results in the conclusion section.
\section{Problem Formulation}
\indent The LBCVRP is defined as follows. Let $G=(V,E)$ be a graph with the set of $n$ vertices $V$, and the set of edges $E$, and let $r \in V$ be the single depot. Each edge $(v_i,v_j) \in E$ has a non-negative weight $c_{ij}$ that represents its length/cost. The edge weights are symmetric (i.e., for each $(v_i,v_j) \in E$, $c_{ij}=c_{ji}$), and satisfy the triangle inequality. Clients are located at the vertices of the graph, and a client at $v \in V \backslash \{ r\} $ has a positive demand $d_v \in \mathbb{N}$. Here, $\mathbb{N}$ denotes the set of natural numbers. We assume $G$ is a complete graph unless otherwise stated. In this paper, the terms nodes, vertices, and clients have the same meaning.\\
\indent A fixed fleet size of $K$ identical vehicles, located at the depot, has to be routed to supply the clients' demands. Each vehicle tour starts and ends at the depot after visiting a subset of the clients such that each client is visited exactly once (i.e. its demand is satisfied entirely by a single vehicle). For a tour $\tau$, the load $d(\tau)$ is the sum of the demands of the nodes contained in $\tau$, and the cost/length $C(\tau )$ is the sum of the edge weights incident on it. A load of each vehicle tour does not exceed the capacity constraint $Q$. Note that, a lower bound for the number of required vehicles is $\left\lceil \frac{\sum\nolimits_{v \in V}d_v}{Q} \right\rceil$.\\
\indent A solution $S=\{\tau_1,\tau_2,...,\tau_K\}$, of the LBCVRP corresponds to a partition $\{ {M_i}:i \in [K]\}$ which verifies the following relations:
$$M_i \ne \emptyset, \forall i; \bigcup_{1 \le i \le K} M_i= V\backslash\{r\}; M_i \cap M_j= \emptyset,  i\ne j; \sum\nolimits_{v \in M_i} {d_v} \le Q, i \in [K].$$
\noindent We use the term "capacitated partition" to refer to such a partition. Indeed, for each $\tau_i=\{r,v_1,...,v_t,r\}\in S$, we have $M_i=\{v_1,v_2,...,v_t\}$. We say, $S$ allocates the clients $\{v_1,v_2,...,v_t\}$ to the $i^{th}$ vehicle.\\
\indent Let $L_S=(d(\tau_1),d(\tau_2),...,d(\tau_K))$ be the vehicles' loads in $S$, $L_S$ may be described as the "load allocation vector" of $S$. Henceforth, when we refer to a load allocation vector $L$, implicitly we assume there is a solution $S$ whose load allocation vector $L_S$, is equal to $L$: $L_S=L$. \\
\indent The objective of the LBCVRP minimizes the total traveling cost and at the same time balances the vehicles' loads. We need to define a "balanced solution" and a "balanced load allocation". The following section gives some preliminaries regarding the balance criteria. We propose a mathematical model for the LBCVRP in section 2.2.
\subsection{Balance Criteria}
\noindent We use the term equity measure/function to refer to an index value that is calculated for a given load allocation. Range fairness is an accepted formulation of the fairness notion, in the load balancing area (see e.g. \cite{Chen2008,Tsouros}). The equity measure, in this case, is the difference between the largest and the smallest load. Ratio fairness (in the sense of the ratio between the largest and the smallest load) is another equity measure. Yousefikhoshbakht et al. \cite{Yousefikhoshbakht} used the ratio measure in a route balanced capacitated vehicle routing problem. \\
\indent One of the widely accepted properties of the equity measures is the Pigou-Dalton (PD) principle \cite{Cowell}, that is applied when the number and sum of outcomes of the allocations are identical: Let $x$  be an allocated vector (e.g.
 the vehicle loads) and $I(x)$ be the equity function. Let $x'$  be formed as follows: ${x'_i} = {x_i} + \delta $ , ${x'_j} = {x_j} - \delta $,  ${x'_k} = {x_k}$ for all $k \notin \{ i,j\} $. The weak PD principle expresses that if $0 \le \delta  \le {x_j} - {x_i}$, then $I(x') \le I(x)$. In the strong version, the inequality is strict ($I(x') < I(x)$) i.e. the new allocation should be more equitable. \\
\indent A simple calculation shows that either the range or the ratio criteria satisfy the weak PD principle. In this paper, we use the ratio measure to evaluate the fairness/balance of the allocated loads. The range measure could be used to obtain similar results. \\
\indent Given the number of vehicles $K$, and a real parameter $0 \le \alpha  \le 1$, a partition of vertices of $V\backslash \{r\}$ into $K$ parts $M=\{ {M_i}:i \in [K]\} $ is said to be $\alpha$-balanced (in the ratio sense) if and only if for each $i,j \in [K]$:
$$\begin{array}{*{20}{c}}
{1 - \alpha  \le \frac{Load({M_i})}{Load({M_j})} \le 1 + \alpha ,\,\,\,\,\,\,\,\,\,\,\,\,\,\,\,\,\,\,\,\,\,\,\,}
\end{array}$$
\noindent where $Load({M_i}) = \sum\nolimits_{v \in {M_i}} {{d_v}}$. This condition simply implies that: 
$$\frac{\max \{ Load(M_i):i \in [K]\}}{\min \{ Load(M_i):i \in [K]\}}  \le 1 + \alpha. \hspace{0.5 cm} (1)$$
\indent We use the "balanced condition" to refer to the inequality (1). It is evident that if a partition is $\alpha_1-$balanced, it is also ${\alpha _2}-$balanced for each ${\alpha _2} \ge {\alpha _1}$. The minimum $\alpha$ for which this condition holds is called the "balanced ratio" of the set $M$. So, 
$$\frac{\max \{ Load(M_i):i \in [K]\}}{\min \{ Load(M_i):i \in [K]\}}-1,$$ 
\noindent is the balanced ratio of the set $M$. 
\begin{definition}Let $S$ be a solution for the LBCVRP with the corresponding partition $M$. We say, $S$ is the balanced solution, iff the balanced ratio of $M$ is the smallest among the balanced ratios of all the capacitated partitions of $V\backslash \{r\}$ into $K$ parts. Also, $S$ is $\alpha-$balanced, iff $M$ is $\alpha-balanced$. The balanced ratio of $S$ is defined to be the balanced ratio of $M$. The load allocation vector of a balanced solution is a balanced load allocation. 
\end{definition}

 \indent For a given load deviation of $\gamma$, $0 \le \gamma  \le Q$ , a simple argument shows that if $\alpha  \le \gamma /Q$  is chosen and the set  $\{ {M_i}:i \in [K]\} $ is $\alpha$-balanced (in the ratio sense), then it is also balanced (in the range sense) with deviation $\gamma$, that is, $\max \{ Load(M_i):i \in [K]\}-{\min \{ Load(M_i):i \in [K]\}} \le \gamma$. \\
 \\
 \indent In this paper, the notation $S$ is used to describe a set of solution for the problem under consideration, $L$ denotes a load allocation vector, $\tau$ denotes a vehicle tour, $\alpha$ is the balanced ratio of a solution, $C$ is used to denote the traveling cost, $K$ is the number of available vehicles, and $T$ is a tree.
\subsection {Mathematical Model}
\indent We assume the depot corresponds to the vertex $0$ and $V= \{1,...,n\}$ represents the set of clients. Moreover, we assign the demand $d_{0}=0$ to the depot. Let $x_{ij}^k$ be the binary variables on the edges $(i,j) \in E$ that decide whether the edge $(i,j)$ is presented in the route of the $k^{th}$ vehicle. An edge $(i,j)$ is presented in the route of the $k^{th}$ vehicle if $x_{ij}^k = 1$ and is not presented otherwise. The following is an integer programming formulation of the LBCVRP:\\
$$\begin{array}{*{20}{c}}
{{C_{opt}} = \mathop {\min \,} \,\sum \limits_{i,j,k}{c_{ij}^{}x_{ij}^k} }&{}&{(2)}\\
\\
{\sum\nolimits_{j = 1}^n {x_{0j}^k}  = 1 }&{\forall k = 1,...,K,}&{(3)}\\
\\
{\sum\nolimits_{i = 0}^n {x_{ij}^k}  - \sum\nolimits_{i = 0}^n {x_{ji}^k = 0} }&{\forall j = 1,...,n, \forall k = 1,...,K,}&{(4)}\\
\\
{\sum\nolimits_{k = 1}^K {\sum\nolimits_{i = 0}^n {x_{ij}^k}  = 1} }&{\forall \,j = 1,...,n,}&{(5)}\\
\\
{0 \le \sum\nolimits_{i = 0}^n {\sum\nolimits_{j = 1}^n {x_{ij}^k{d_j}} }  \le Q}&{\forall k=1,...,K,}&{(6)}\\
\\
{1 - \alpha  \le \frac{\sum\nolimits_{i = 0}^n {\sum\nolimits_{j = 1}^n {x_{ij}^k{d_j}} } }{\sum\nolimits_{i = 0}^n {\sum\nolimits_{j = 1}^n {x_{ij}^l{d_j}} }} \le 1 + \alpha }&{\forall k,l=1,...,K,}&{(7)}\\
\\
{x_{ij}^k \in \{ 0,1\} }&{\forall i,j = 0,...,n, \forall k=1,...,K.}&{(8)}
\end{array}$$\\
\indent  Equality equations (3) ensure that all the vehicles depart from the depot. The constraints (4) indicate that a vehicle will depart from a client after visiting it. The constraints (5) specify the fact that each client is visited exactly once. Constraints (6) indicate that a vehicle cannot carry more than its capacity Q. The balance restriction for loads of the vehicles is guaranteed by the constraints (7). Restrictions (8) indicate that each edge in the graph has the value 1, if it is used, and 0 otherwise. \\
\indent  By the above formulation, the LBCVRP is programmed as the basic CVRP with the extra constraints of the balanced condition (constraints 7). To avoid ambiguity, we call this problem $\alpha {\rm{ - LBCVRP}}$, when $\alpha$ is fixed. \\
\indent Using the equations (2)-(8), the inequalities (7) impose feasibility on the solutions. We will show finding a feasible solution for the $\alpha {\rm{ - LBCVRP}}$, that satisfies the related balanced condition, is NP-complete (see sub-section 2.3). Let us consider an instance $I$ of the LBCVRP where the distances between the locations are identical, and the clients' demands are as indicated in Fig. 1. There are a fleet of 2 identical vehicles in the depot, each having a capacity of 5. All possible solutions are listed below: ${S_1} = \{ \{ r,1,2,r\} ,\{ r,4,r\} \} $, ${S_2} = \{ \{ r,1,4,r\} ,\{ r,2,r\} \}$. The balanced ratio of the set ${S_1}$ is $\alpha _1 = 1/3$, and the balanced ratio of $S_2$ is $\alpha _2 = 2/3$. The minimum ratio is related to ${S_1}$, and there is no solution with a balanced ratio smaller than $1/3$. So, $S_1$ is a balanced solution. \\
\indent One of the major problems is to find the lowest value of $\alpha$ for which the $\alpha {\rm{ - LBCVRP}}$ has a feasible solution. We show a decision version ${I_\alpha }$ of this problem that asks to find whether it has a feasible solution, is NP-complete (see section 2.3), so determining the minimum value of $\alpha$ is NP-hard. \\
\begin{figure}[ht]
\begin{center}
\includegraphics[scale=0.28] {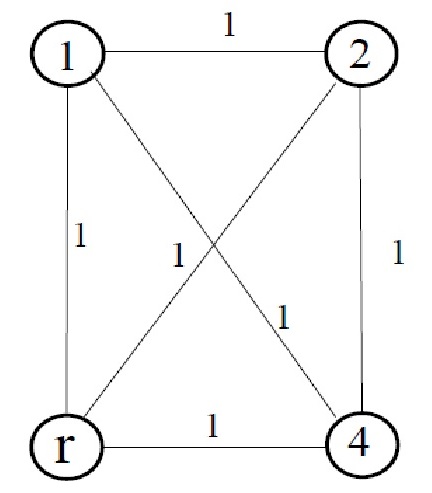} 
\end{center}
\caption{An instance of the LBCVRP that has no solution with a balanced ratio smaller than 1/3.}
\label{fig1}
\end{figure}
\indent  The LBCVRP could be analyzed from another viewpoint as a multi-objective problem (see \cite{Bowerman}). This approach could be a good way to round around the difficulties of the feasibility. There are two objectives:\\
 \noindent $P_1$: Minimization of the total traveling cost.\\
 \noindent $P_2$: Minimization of the function $\sum\nolimits_{k = 1}^K {(\sum\nolimits_{i = 0}^n {\sum\nolimits_{j = 1}^n {x_{ij}^k{d_j}} } )^2} $.\\
 \indent The objective function ${P_2}$ has been proposed by Bowerman et al. \cite{Bowerman}. We define the objective function $P= \lambda {P_1} + (1 - \lambda ){P_2}$, for $0 \le \lambda  \le 1$. The parameter weightings effects can be examined by decision-makers to find the desired set of vehicle tours. For example, if load balancing introduces costly tours, then one can reduce the emphasis on load balancing. Similarly, if a desirable solution cannot be reached by the cost measure, then one can increase the weighting of the cost measure to force the tours to become inexpensive. We present a 4-approximation algorithm for the LBCVRP with the aforementioned objective function. 
\subsection{Complexity Analysis}
\noindent Here, we show the ${\rm{\alpha-LBCVRP}}$ is NP-hard. We need to prove that its decision version is NP-complete (see \cite{Garey}). Let ${I_\alpha }$  for $0 \le \alpha \le 1$ be a decision version of this problem that asks to find whether it has a feasible solution. The following theorem indicates ${I_\alpha }$  is NP-complete for each $\alpha$, $0 \le \alpha \le 1$. 
\begin{theorem}
For each $\alpha$, $0 \le \alpha \le 1$, ${I_\alpha }$  is an NP-complete problem.
\end{theorem}
\begin{proof}
\noindent  First, we show ${I_0 }$ is NP-complete by a reduction to the partition problem. The partition problem is the task of deciding whether a given set $A = \{ {a_1},{a_2},...,{a_n}\}$ of positive integers has a sub-set $A' \subseteq A$  such that $\sum\nolimits_{{a_i} \in A'} {{a_i}}  = \sum\nolimits_{{a_j} \in A\backslash A'} {{a_j}} $. This problem is a known NP-complete problem. We define $V = \{ {v_1},{v_2},...,{v_n},r\} $ to be the set of clients and the depot. For each ${v_i} \in V\backslash \{ r\} $, we assign the demand ${d_{{v_i}}} = {a_i}$. Let the distances between the clients and the depot be the same and the depot includes two vehicles each of capacity $Q$, $\sum\nolimits_{i = 1}^n {{a_i}} /2 \le Q < \sum\nolimits_{i = 1}^n {{a_i}} $. This problem is an instance of ${I_0 }$ and has a feasible solution, iff the partition problem has a solution. Furthermore, this reduction can be done in polynomial time. Hence, the ${I_0 }$ is an NP-complete problem. \\
\indent Now, we show $I_\alpha$ is NP-complete for each $\alpha$, $0 < \alpha  \le 1$. We reduce $I_{\alpha}$ to the $I_0$ in polynomial time. Let $I'_{I_0}$ be a given instance of the $I_{0}$ that has a set of clients $ \{v_1,v_2,...,v_n\}$ with demands $\{d_1,d_2,...,d_n\}$. The number of vehicles in the depot is $K_{I'_{I_0}}$, and the capacity constraint is $Q_{I'_{I_0}}$. We construct an instance $I'_{I_\alpha }$ of $I_\alpha$ similar to $I'_{I_0}$, and just restrict the vehicle capacity constraint to $Q_{I'_{I_\alpha}}=\sum\nolimits_{i = 1}^n {d_i} /K_{I'_{I_0}}$. Clearly, $x$ is a feasible solution of $I'_{I_\alpha}$, iff it is a feasible solution of $I'_{I_0}$. This proves that for each $0 \le \alpha \le 1$, ${I_\alpha }$ is NP-complete.\\
 $\qed$
\end{proof}

\indent We conclude (from theorem 1) that finding the smallest value of $\alpha$ for the LBCVRP is NP-hard. In Section 3, we give an algorithm called "load balancing algorithm" that finds (in polynomial time) the smallest value of $\alpha$ for the LBECVRP. However, the LBECVRP is still NP-hard, since it is a generalization of the ECVRP. The complexity of solving the ECVRP relies on the capacity of the vehicles. When the vehicle capacity is 2, the ECVRP can be solved in polynomial time \cite{Asano1996}. However, it is NP-hard for any $Q\ge3$ \cite{Asano1996}. Albeit, the authors in \cite{Asano1996} showed that the ECVRP is indeed APX-complete for any $Q\ge 3$, i.e., there exists $\epsilon> 0$ such that no $1 + \epsilon$ approximation algorithm exists unless $P = NP$.
 \section{LBECVRP}
\noindent In this section, we consider the LBECVRP and show that the fairest solution can be found. We present two approximation algorithms for the LBECVRP in general metrics and on trees that find fairest solutions. First, we define what we mean by the fairest solution, and then explain how we can find such a solution.\\
\indent We seek a set of vehicle tours that start from the depot and terminate there after visiting a set of clients. So, a routing has two components: the traveling route of a tour and the allocation of the clients that the tour will visit. The allocation of the clients should be as fair as possible (in the ratio sense). \\
\indent Let $L=(L_1,L_2,...,L_K)$ be a load allocation vector, that is sorted in non-decreasing order, we define $R_{L}=(\frac{L_K}{L_1},...,\frac{L_2}{L_1},1)$. We say $L$ is the fairest (in the ratio sense) if the $K-$tuple $R_L$ is the fairest in the min-max sense. Intuitively, the min-max fairness can be derived from the following approach: first, make sure that the maximum ratio is as small as possible and then ignoring this maximum ratio, make sure that the ratio between the load of the vehicles that can still get additional load and the minimum load is the smallest, and so on.\\
\indent Moreprecisly, let $R= ({w_1},{w_2},...,{w_K})$ and $R' = ({w'_1},{w'_2},...,{w'_K})$ be two $K-$tuples, each sorted in non-increasing order. We say $R'$ is lexicographically smaller than $R$, if $R=R'$ or there is some index $j$ for which ${w'_j}<{w_j}$ and ${w'_i}= {w_i}$ for all $i<j$. We define a partial order (denoted by $\prec$) on the set including all the load allocation vectors. Given two load allocation vectors $L$ and $L'$, we say $L'$ is fair than $L$ (written $L' \prec L$), iff $R_{L'}$ is lexicographically smaller than $R_L$. We will say $L$ and $L'$ are equivalent, iff both $L' \prec L$ and $L \prec L'$. So, $L$ and $L'$ are equivalent, iff $R_L=R_{L'}$. Thus, the relation $\prec$ is a total order on the equivalence classes of the load vectors. The fairest allocations are the vectors in the unique minimal equivalence class under $\prec$. 
\begin{definition} Let $S$ be a solution for the LBECVRP, and $L_S$ be the allocated loads. We say $S$ is the fairest solution, iff $L_S$ is a fairest load allocation vector, that is, $L_S$ belongs to the unique minimal equivalence class under $\prec$.
\end{definition} 

\indent Definition 2, brings the relationship between the balanced solution and the fairest solution. The fairest solution has the smallest balanced ratio among the balanced ratios of all the capacitated partitions of the vertices of $V\backslash \{r\}$ into $K$ parts. So, the fairest solution is indeed the balanced solution. However, the balanced solution may not be the fairest. In the following section, we present an algorithm called "load balancing algorithm" that explains how we can find such a fairest solution.  
\subsection{Load Balancing Algorithm}
\noindent In this section, we consider the following load balancing problem. Let $I = \{ {i_1},...,{i_n}\}$ be a set of items each with volume of $1$, and $B = \{ {B_1},...,{B_K}\} $ be a set of bins. We assume each item $i_j$, $j=1,...,n$ can be put in any bin in the set $B$. All bins have the same capacity, denoted by $Q$. We wish to put each item in a bin, and our optimum solution $F^* = ({L^*_1},...,{L^*_K})$  specifies the number of items (total volume) each bin $B_i$ receives. A packing is a function $F: I \to B$ that puts each item $i_j$ in a bin in $B$. Note that, the items correspond to the clients, and the bins relate to the vehicles. So, the optimum solution $F^*$ is indeed a load allocation to the $K$ vehicles. Algorithm 1 finds the optimum packing for this problem. Lemma 1 shows this packing is indeed the best possible.\\
\indent More generally, each item $i_j$ can have a different volume $d_j\ge 1$: each item should be put in a unique bin. This problem can be encoded in the un-splittable assignment problem (see \cite{Lenstra} for a detailed definition of the un-splittable assignment problem). \\

\begin{algorithm}

\caption{Load balancing algorithm}

$\mathbf{input:}$ A set of $n$ identical items that should be packed into a set of $K$ bins.\\
$\mathbf{output:}$ Optimum packing $F^*$.
\begin{itemize}
\item[] $\mathbf{Start}$ 
\item[] Designate $\left\lfloor {\frac{n}{K}} \right\rfloor $ items to each bin.

\item[] Distribute the rest items to $n-\left\lfloor {\frac{n}{K}} \right\rfloor$ bins randomly. 

\item[] Sort the allocated loads in non-decreasing order to find the optimum packing $F^* = ({L^*_1},...,{L^*_K})$.
\item[] $\mathbf{End}$
\end{itemize}

\end{algorithm}

\indent For example, assume there are $n=10$ items and $K=3$ bins each with a capacity of $Q=5$. The best packing found using the algorithm is $F^*=(3,3,4)$; another packing $\bar{F}$ may be e.g. $\bar{F}=(2,3,5)$. In a packing $F$, let $b_i$ be the number of packed items into the bin $B_i$, we will also refer to $b_i$ as the degree of $B_i$. Let $l^*$ be the maximum degree of the optimum packing $F^*$. In the above example, we have $l^*=4$, also the value of the maximum degree of $\bar{F}$ is $5$. Lemma 1 shows that the packing $F^*$ is indeed the best possible. 
\begin{lemma}The load balancing algorithm finds an optimum packing of the items to the bins. 
\end{lemma}
\begin{proof} Let  ${F^*} = (L_1^*,L_2^*,...,L_K^*)$ be the optimum packing found by the algorithm, and $F' = ({L'_1},{L'_2},...,{L'_K})$ be some other packing, each sorted in non-decreasing order. The packings $F^*$ and $F'$ are indeed two different load allocation vectors. Let ${R_{F^*}} = (\frac{L_K^*}{L_1^*},...,\frac{L_2^*}{L_1^*},1)$ and $R_{F'}= (\frac{L'_K}{L'_1},...,\frac{L'_2}{L'_1},1)$, each are sorted in non-increasing order. We see the entries in ${F^*}$ are equal to ${l^*}$ or ${l^*-1}$. Let $b^*$ be the number of bins with degree ${l^*}$. If ${b^*}=K$, obviously the packing ${F^*}$ is the best possible. So, we assume that ${b^*}<K$.\\

\noindent $\mathbf{Claim\,\,1} $ ${L'_K} \ge {l^*}$.
\begin{proof}Otherwise, let ${L'_K} < {l^*}$, so we have $n = \sum\nolimits_{i = 1}^K {L'_i}  \le K({l^*} - 1)$. This is in contradiction with the fact that: $n = \sum\nolimits_{i = 1}^K {L_i^*}  = {b^*}{l^*} + (K - {b^*})({l^*} - 1)$ and ${{b^*}} \ge 1$.
\end{proof}

\indent When the largest entry $L'_K$ of $F'$ is equal to $l^*$, the smallest entry $L'_1$ is at most $l^*- 1$, so we have $\frac{L'_K}{L'_1} \ge \frac{L_K^*}{L_1^*}$. The equality  $\frac{L'_K}{L'_1} = \frac{L_K^*}{L_1^*}$ occurs when $F^* = F'$. If ${L'_K} > {l^*}$, there exists an index $1 < h \le K$ for which $L_h^* > L'_h$ because of $\sum\nolimits_{i = 1}^K {L_i^*}  = \sum\nolimits_{i = 1}^K {L'_i}  = n$. As it is stated before, the entries in $F^*$ are equal to $l^*$ or $l^*-1$, so we have either $L_h^*=l^*$, or $L_h^*=l^*-1$. It is easy to see that both the cases imply: $\frac{L'_K}{L'_1} >\frac{ L_K^*}{L_1^*}$. This proves that the $K-$tuple $R_{F^*} = (\frac{L_K^*}{L_1^*},...,\frac{L_2^*}{L_1^*},1)$ is lexicographically smaller than $R_{F'}= (\frac{L'_K}{L'_1},...,\frac{L'_2}{L'_1},1)$. Thus, $F^*=(L_1^*,L_2^*,...,L_K^*)$ is a fairest load allocation vector.\\
 $\qed$
\end{proof}

\indent  We assume $L_1^*, L_2^*, ..., L_K^*$  is the sequence of the vehicles' loads found using the load balancing algorithm. There are two possible cases: $L_K^*=L_1^*$ or $L_K^*=L_1^*+1$. In the following, we assume $L_K^* \ge 3$, and endeavor to find a set of tours whose loads are equal to $F^*$. 
\subsection{LBECVRP in general metrics}
\noindent Our algorithm employs a tour partitioning heuristic (see \cite{Beasley}) to find the fairest solution (see algorithm 2). It has a similar flavor as the heuristics given in \cite{Altinkemer1990,Altinkemer1987,Haimovich1985}.  \\

\begin{algorithm}
\caption {} 
$\mathbf{input:}$ An instance of the LBECVRP together optimum loads $L_1^*,...,L_K^*$.\\
$\mathbf{output:}$ A set of feasible vehicle tours whose loads are balanced.
\begin{itemize}
\item[] $\mathbf{Start}$
\item[] Use the Christofides algorithm \cite{Christofides} to find a salesman tour $\tau$  spanning the vertices in $V$.  Let $p: = ({v_1},{v_2},...,{v_n})$ be the vertices in the order of their representation in $\tau$. 
\item[] $\mathbf{for}$ $i= 1$ to $ L_1^*$ $\mathbf{do}$
\begin{itemize}
\item[] Begin at $v_i$ and find the following tours:\\
 $S_i=\{\tau _1^i = (r,{v_i},{v_{i+1}},...,{v_{i+L_K^*-1}},r), \tau _2^i = (r,{v_{i+L_K^* }},{v_{i+L_K^* + 1}},...,{v_{i+L_K^* + L_{K - 1}^*-1}},r),...,\\$ $\tau _K^i = (r,{v_{i-L_1^*}},...,{v_{i-1}},r)\}$. 
\item[] Find the total traveling cost of $S_i$.
\end{itemize}
\item[] $\mathbf{end\,\, for}$
\item[]  Return the solution ${S_H}: = \{ \tau _1^H,\tau _2^H,...,\tau _K^H\} $,  $1 \le H \le L_1^*$ whose total traveling cost is the smallest. 
\item[] $\mathbf{End}$
\end{itemize}
\end{algorithm}
\indent In the following, we prove that the traveling cost of $S_H$ is within $3-\frac{3}{2Q}$ times the optimal value $C_{LBECVRP}$ of the LBECVRP. First, we prove a lemma to give a lower bound for the $C_{LBECVRP}$. Here, $c_{rv}$ is the edge cost connecting $r$ to $v$ for a node $v \in V$.
\begin{lemma}
${C_{LBECVRP}} \ge 2\frac{{\sum\nolimits_{v \in V}^{} d_v c_{rv}}}{L_K^*}.$
\end{lemma}
\begin{proof}
\indent Let $S_{opt}: = \{ \tau _1,\tau _2,...,\tau _K\}$ be an optimal solution for the LBECVRP and $\tau_i \in S_{opt}$. Let $c_{rv}^{max} = \max \{c_{rv}:v \in {\tau_i}\} $ and $C({\tau_i})$ be the traveling cost of $\tau_i$. Since $\sum\nolimits_{v \in \tau_i} d_v \le L_K^*$, we have:
$$C(\tau_i) \ge 2c_{rv}^{max} = 2\frac{{\sum\nolimits_{v \in \tau_i} d_v }}{{\sum\nolimits_{v \in \tau_i} d_v }}c_{rv}^{max} \ge 2\frac{{\sum\nolimits_{v \in \tau_i} d_v }}{L_K^*} c_{rv}^{max} \ge 2\frac{\sum\nolimits_{v \in {\tau_i}} d_v c_{rv} }{L_K^*}.\hspace{2cm}$$
\noindent Summing over all the tours in $S_{opt}$, we obtain:
$${C_{LBECVRP}}=\sum\nolimits_{i=1}^{i=K} C( \tau_i) \ge 2\frac{\sum\nolimits_{v \in V}{d_v c_{rv}} }{L_K^*}. $$
$\qed$
\end{proof}
\begin{theorem}
The approximation factor of the algorithm 2 for the LBECVRP is $3-\frac{3}{2Q}$.
\end{theorem}
\begin{proof}
Let $C(\tau)$ be the cost of the traveling salesman tour $\tau$ covering the vertices in $V$. Since the Christofides algorithm is used to obtain $\tau$, we have $C(\tau ) \le \frac{3}{2} C_{LBECVRP}$. Indeed, the cost of an optimal traveling salesman tour is a lower bound for the $C_{LBECVRP}$. We see that each node $v\in V \backslash \{r\}$ appears at most once as the first node of a tour and at most once as the end node of a tour during $L_1^*$ iterations of the partitioning procedure. Hence, each edge $(r,v)$ appears at most 2 times during the iterations. Besides, the edge $(u,v) \in \tau$ is not included in the solution where $v$ appears as the first node of a tour. Thus,
$$\sum\nolimits_{j=1}^{L^*_1}\sum\nolimits_{i = 1}^{K} {C({\tau^j_i})}  \le (L_1^*-1)C(\tau ) + 2\sum\nolimits_{v \in V} c_{rv}. \hspace{0.5 cm} (9)$$
\noindent Note that, if $v$ does not appear as the first node of any tour during the $L_1^*$ iterations of the partitioning procedure, we then have: $c_{uv} \le c_{ru}+c_{rv}$ due to the triangle inequality. Hence, the right-hand side of (9) is an upper bound for the total traveling cost. Since $S_H$ has a minimum traveling cost among the others, we have:
$$L_1^* C(S_H)\le\sum\nolimits_{j=1}^{L^*_1}\sum\nolimits_{i = 1}^{K} {C({\tau^j_i})}  \le (L_1^*-1)C(\tau ) + 2\sum\nolimits_{v \in V} c_{rv}. $$
\noindent Besides, according to the load balancing algorithm, there are two cases: (a) $L_K^*=L_1^*$, (b) $L_K^*=L_1^*+1$.\\
\indent  First, assume $L_K^*=L_1^*$. Hence,
$$\begin{array}{l}
C(S_H) \le  (1-\frac{1}{L^*_1})C(\tau ) + \frac{2\sum\nolimits_{v \in V} c_{rv}}{L_1^*} \\
\hspace{1.13cm}\le  (1-\frac{1}{Q})C(\tau ) + \frac{2\sum\nolimits_{v \in V} c_{rv}}{L_K^*} \\
\hspace{1.13cm}\le  ((1-\frac{1}{Q})\rho  + 1){C_{LBECVRP}} = (2.5-\frac{3}{2Q}) C_{LBECVRP}.
\end{array}$$
\noindent The second inequality follows from the fact that $L_1^* \le Q$. The third inequality follows from lemma 2. 

\indent Now, let $L_K^*=L_1^*+1$. Since $L^*_K \ge 3$, we have $\frac{2\sum\nolimits_{v \in V} c_{rv}}{L_1^*}\le \frac{3\sum\nolimits_{v \in V} c_{rv}}{L_K^*}$. Hence,
$$\begin{array}{l}
C(S_H) \le  (1-\frac{1}{L^*_1})C(\tau ) + \frac{2\sum\nolimits_{v \in V} c_{rv}}{L_1^*} \\
\hspace{1.13cm}\le  (1-\frac{1}{Q})C(\tau ) + \frac{3\sum\nolimits_{v \in V} c_{rv}}{L_K^*} \\
\hspace{1.13cm}\le  ((1-\frac{1}{Q})\rho  + \frac{3}{2}){C_{LBECVRP}} = (3-\frac{3}{2Q}) C_{LBECVRP}.
\end{array}$$
\noindent Here, $\rho$ is the factor of approximation for the TSP.\\
 $\qed$
\end{proof}
\noindent $\mathbf{Remark}$ When the metric space is a tree, TSP can be solved optimally. Thus, the approximation factor of the above algorithm is $2.5-\frac{1}{Q}$ for tree networks, when $L_K^*=L_1^*+1$. In the following, we present an improved $2$-approximation algorithm for the TLBECVRP.
\subsection{LBECVRP on Trees}
\noindent In this section, we present a $2-$approximation algorithm for the TLBECVRP. In this case, a vehicle tour can be characterized by a rooted subtree, and so more than one vehicle may pass through the nodes of the tree. \\
\indent  Let $T=(V,E)$ be a tree with the set of vertices $V$ and the set of edges $E$. We assume $T$ is binary (as otherwise, we can split high degree vertices, and add zero-length edges), and rooted at the depot $r$. We denote the unique path between the vertices $u,v \in V$ by $path(u,v)$, and identify its cost by $C(path(u,v))$. For any $v \in V$, the cost $C(T_v)$ is the sum of the edge weights in $T_v$. Here, $T_v$ is the subtree rooted at $v$. The load $d(T_v)$ is defined similarly: $d(T_v)=\sum_{v_i \in T_v}d_{v_i}$. A solution to the TLBECVRP consists of a set of tours. We specify a tour by a set of clients that the vehicle visits. In fact, for a given set $V_k \subseteq V \backslash \{ r\}$, an optimal tour for $V_k$ can be trivially obtained: first compute a minimal subtree that spans $V_k \cup \{ r\}$, and then perform a depth-first search starting at vertex $r$. Thus, we describe the $k^{th}$ vehicle tour by ${\tau_k}= \{ v|v\,\,is\,\,served\,\,by\,\,the\,k^{th}\,\,vehicle\}$, considering the fact that all the demands are equal to $1$. \\
\indent Assume $L^*_1, L^*_2, ..., L_K^*$ is the sequence of the fairest loads and $L^*_K= L^*_1+1$. We define a lower bound for the edge $e=(u,v) \in T$ in an optimal solution as follows:
$$LB(e) = 2c_{uv}.\left\lceil {\frac{d(T_v)}{L^*_K}} \right\rceil.$$
\indent $LB(e)$ is a lower bound for the edge $e$, since at least $\left\lceil {\frac{d(T_v)}{L^*_K}} \right\rceil$ number of vehicles is required (for any solution) to serve the clients in $T_v$, and each vehicle traverses $e$ at least twice. In the following lemma, we give a lower bound for the optimal cost of the TLBECVRP.
\begin{lemma}
$LB =\sum _{e \in E} LB(e)$, is a lower bound for the optimal cost of the TLBECVRP.
\end{lemma}

\indent Our algorithm (see algorithm 3) is similar to the algorithms given in \cite{Asano2001,Hamaguchi,Naoki}. It proceeds in a sequence of rounds. Let $K'$ be the number of loads that are equal to $L^*_K$, in the fairest load allocation. In the first $K'$ rounds, the algorithm finds some tours whose loads are equal to $L_K^*$. In each round, it focuses on a particular $L_K^*$-feasible node $u$ which is $L_K^*$-minimal and proposes a strategy. A vertex $v \in V$ is known as $L_K^*-$feasible if $d(T_v) \ge L_K^*$, and is known as $L_K^*-$minimal, if it is $L_K^*$-feasible but none of its children is. In the next rounds, it finds some tours whose loads are equal to $L_1^*$. In these rounds, we reform the tree to reason better about the structure of the resulting instance. We use the reforming operations given in \cite{Asano2001} and modify them ensuring that they preserve the lower bound and do not decrease the optimum cost. These operations can be made safely at any point in the algorithm until one of a few cases arise. Each case has a corresponding strategy that finds a set of tours. A feasible solution in the modified tree has a corresponding feasible solution in the original tree. In the following, we briefly restate the reforming operations given by Asano et al. \cite{Asano2001} and refer the reader to that paper for more details.\\
 \indent $\mathbf{Reforming \,\,Operations.}$ We assume that the vertices with positive demands are exactly the leaves: if some internal vertex $v$ has positive demand $d_v=1$, we add a vertex $v'$ with demand $d_{v'}=1$ and edge $(v,v')$ of length zero and set $d_v$ to zero. Besides, we assume no non-root vertex has a degree exactly two, as no branching would occur there, and the two incident edges can be spliced into one. In the following, we list the reforming operations. Here, we use the names given in \cite{Becker}. 
\begin{itemize} 
\item[] $\mathbf{Condense:}$ The condense operation merges a subtree whose demand is at most $L_K^*$ into an edge. Precisely, for an internal node $v$ with parent  $u$, and $d(T_v) \le L_K^*$, the condense operation replaces $T_v$ by an edge $(v,v')$ whose weight is $c_{vv'}=C(T_v)$ and $d_{v'}=d(T_v)$. Then, the vertex $v$ has a degree exactly two, and the two incident edges can be spliced into one that has the weight $c_{uv'}=c_{uv}+c_{vv'}$. If $d_{v'}=L_K^*$, then we delete the edge $(u,v')$ and add the edge $(r,v')$ with the cost $c_{rv'}=C(path(r,u))+c_{uv'}$. The demand of $d_{v'}$ will be considered in the final round of the algorithm. 
\item[] $\mathbf{Unit:}$ The unit operation merges leaves of nodes. Let $\{v_1,v_2,...,v_k\}$ be a subset of leaves of a node $u$. Here, $c_{uv_i}$ denotes the weight of the edge $(u,v_i)$. The unit operation examines any pair of leaves $v_i, v_j$ and merges them, if $d_{v_i}+d_{v_j} \le L_K^*$. Exactly speaking, it replaces the demand of $v_i$ with $d_{v_i}=d_{v_i}+d_{v_j}$ and the weight $c_{uv_i}$ with $c_{uv_i}+c_{uv_j}$ and then removes the leaf $v_j$ together with the edge $(u,v_j)$. If $d_{v_i}=L_K^*$, then we delete the edge $(u,v_i)$ and add the edge $(r,v_i)$ with the cost $c_{rv_i}=C(path(r,u))+c_{uv_i}$. The demands of $d_{v_i}$ will be considered in the final round of the algorithm. 
\item[] $\mathbf{Unzip:}$ If a node $v$ with parent  $u$ has two children $v_1,v_2$ and $L_K^* <d(T_{v_1})+d(T_{v_2}) < 2L_K^*$, then the unzip operation deletes $v$ and adds edges $(u,v_1), (u, v_2)$ with weights $c_{uv_i}=c_{uv}+c_{vv_i}$ for $i=1,2$.

\indent Asano et al. \cite{Asano2001} introduced the notion of a p-node and q-node. A p-node is an internal node whose children are leaves, and the sum of their demands is between $L_K^*$ and $2L_K^*$. A q-node $u$ is an internal node for which the load of the subtree $T_u$ (rooted at $u$) is at least $2L_K^*$, but none of the children of $u$ has this property. Note that, the demands at the leaves of a q-node are smaller than $L_K^*$.
\item[] $\mathbf{Merge:}$ If a node $u$ has a child of p-node $v$, and some other leaves directly connected to $u$, and for a leaf node $v_i$, $d_{T_v}+d_{v_i}<2L_K^*$, then $v_i$ is placed as the child of $v$ with the edge of weight equal to $c_{uv_i}$. 
\end{itemize}

\indent  Performing the unit, unzip and merge operations ensures that a p-node has exactly three leaves. If it has four leaves, then their demands exceed $2L_K^*$ which is in contradiction with the definition of a p-node. If it has two children, then the unzip operation connects its children directly by its parent. If it has one child, then its degree is exactly two, so the two incident edges can be spliced into one. It is obvious that the lower bound is not increased by the above reformations. However, the upper bound increases, since the possible tours are restricted. There are three possible cases for a q-node, after the reforming operations (see Fig. 2). 
\begin{itemize}
\item[] $\mathbf{Case 1.}$ There exists more than one child of p-node in a q-node.
\item[] $\mathbf{Case 2.}$ There exists no child of p-node in a q-node.
\item[] $\mathbf{Case 3.}$ There exists only one child of p-node in a q-node.\\
\end{itemize}
 
\begin{figure}[ht]
\begin{center} 
\includegraphics [scale=0.38]{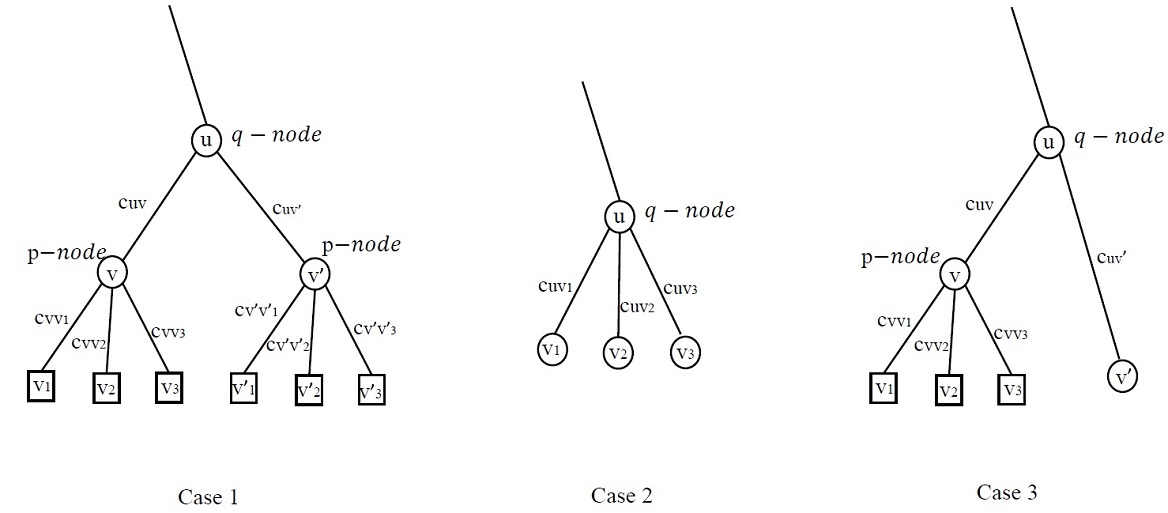} 
\end{center}
\caption{Three possible cases for a q-node.}
\label{fig2}
\end{figure}
\indent Our algorithm performs in a sequence of rounds similar to that of given by Asano et al. \cite{Asano2001} and Hamaguchi and Katoh \cite{Hamaguchi}. The main difference between our algorithm and their algorithms is that we need loads of the vehicles to be exactly equal to $L_K^*$ or $L_1^*$ (our given bounds), and the number of loads that are equal to $L_K^*$ is limited, while in their algorithms the number of vehicles is not limited and the vehicles' loads are at the peak of the given bound $Q$ (not exactly $Q$).\\
\indent  In each round, we apply an appropriate strategy to partition the tree. We need to compute the cost of the tours needed by the strategy and the reduced quantity from the lower bound. Let $I$ be the problem instance for which we will apply the strategy, and $LB(I)$ be the lower bound of its optimal cost given in lemma 3. Let $I'$ be the problem instance obtained after applying the strategy, and $LB(I')$ be the lower bound of its optimal cost. The reduced quantity from the lower bound is defined as $LB(I)- LB(I')$. The final round is when we do not have any q-node. In this round, there are some leaves whose loads are equal to $L_K^*$, and there are either at most two leaves whose loads are smaller than $L_K^*$ or at most one p-node. Since the vehicles' loads are equal to $L_1^*$, the load of the remaining tree is a multiple of $L_1^*$. 

\begin{algorithm}
\caption {} 
$\mathbf{input:}$ An instance of the TLBECVRP together optimum loads $L_1^*,...,L_K^*$.\\
$\mathbf{output:}$ A set of feasible vehicle tours whose loads are balanced.
\begin{itemize}
\item[] $\mathbf{Start}$
\item[]  $K':=$ Number of optimum loads that are equal to $L^*_K$.
\item[] $\mathbf{for}$ $ i=1$ to $ K'$ $\mathbf{do}$
\begin{itemize}
\item[] Apply strategy 1 to find some tours whose loads are equal to $L_K^*$.
\item[] Calculate the cost of the tours needed by the strategy and the reduced quantity from the lower bound. 
\item[] Compute the ratio between the cost of the tours found and the reduced quantity from the lower bound.
\end{itemize}
\item[] $\mathbf{end \,\, for}$
\item[] $\mathbf{for}$ $ i> K'$ $\mathbf{do}$
\begin{itemize}
\item[] Apply the reforming operations to the remaining tree until any of them cannot be applied.
\item[] Focus on a particular q-node $u$ and apply an appropriate strategy among the strategies $2-5$ to find some tours whose loads are equal to $L_1^*$.
\item[] Calculate the cost of the tours needed by each strategy and the reduced quantity from the lower bound. 
\item[] Compute the ratio between the cost of the tours found and the reduced quantity from the lower bound.
\item[] $\mathbf{if}$ i is the final round (there is no q-node) $\mathbf{do}$
\begin{itemize}
\item[] Apply strategy 6 to find a set of tours whose loads are equal to $L_1^*$.
\item[] Calculate the cost of the tours needed by the strategy and the reduced quantity from the lower bound. 
\item[] Compute the ratio between the cost of the tours found and the reduced quantity from the lower bound.
\end{itemize}
\item[] $\mathbf{end\,\, if}$ 
\end{itemize}
\item[] $\mathbf{end \,\,for}$
\item[] Return the set including the feasible tours found.
\item[] $\mathbf{End}$
\end{itemize} 
\end{algorithm} 
\begin{theorem} The approximation factor of the algorithm 3 for the TLBECVRP is $2$.
\end{theorem}
\noindent $\mathbf{Sketch \, of \, proof.}$ The proof is by induction on the number of rounds similar to that of given in \cite{Asano2001,Naoki}. We assume the theorem holds for the problem instances that require at most $K-1$ rounds, and consider the problem instance $I$ for which our algorithm requires $K$ rounds. Let $I'$ be an instance of the TLBECVRP obtained from $I$ after the first round whose lower bound is $LB(I')$, and let $LB_1$ be the reduced quantity from the lower bound at this round. Let $cost(I)$, $cost_1$ and $cost(I')$ denote the total cost that is required by our algorithm to performs on the original problem $I$, the cost of our algorithm in the first round and the total cost that is required by our algorithm to performs on the remaining problem $I'$, respectively, (i.e., $cost(I)=cost_1 + cost(I')$). Then, we have
$$\frac{cost(I)}{LB(I)} \le \frac{cost_1+cost(I')}{LB_1+LB(I')}.$$
\noindent From the induction hypothesis, we have $\frac{cost(I')}{LB(I')} \le 2$, so it suffices to prove
$\frac{cost_1}{LB_1} \le 2.$
\noindent Lemmas 4 through 8 prove that this inequality holds in each case. Thus, we have the theorem. \\
$\qed$\\

\indent Let $k$ be the number of rounds of the algorithm. The proposed algorithm performs the following strategy iteratively for each $k=1,...,K'$. The algorithm finds a deepest vertex $v \in T$ s.t. the load of the subtree $T_{v}$ below $v$ is at least $L_K^*$. So, the load of the subtrees hanging off the $v$'s children are smaller than $L^*_K$. Note that, we have assumed $T$ is binary and rooted at $r$. Let $v_1, v_2$ be the children of $v$ that are connected to $v$ by the edges $(v,v_1)$ and $(v,v_2)$ with weights/costs $c_{vv_1}$ and $c_{vv_2}$. The algorithm considers subtrees $T_{v_1}$ and $T_{v_2}$ satisfying 
$$d(T_{v_1}) < L^*_K,\,\,\,d(T_{v_2}) < L^*_K,\,\,\,d(T_{v_1})+d(T_{v_2})+d_{v} \ge L^*_K.$$
\noindent Without loss of generality, we assume $C(T_{v_1})+c_{vv_1} \ge C(T_{v_2}) + c_{vv_2} $. We split the vertices $V_{T_{v_2}}=\{u: u\in T_{v_2}\}$ of $T_{v_2}$ into $V^1_{T_{v_2}}$ and $V^2_{T_{v_2}}$ so that $d(T_{v_1})+d(V^1_{T_{v_2}})=L^*_1$ where $d(V^1_{T_{v_2}})=\sum_{i\in V^1_{T_{v_2}}} 1$.\\
\indent $\mathbf{Strategy \, 1}:$ We allocate a vehicle to serve the clients in $T_{v_1} \cup \{v\}$ and $V^1_{T_{v_2}}$. The computation of such $V^1_{T_{v_2}}$ is straightforward: perform a depth-first search on $T_{v_2}$ starting at $v_2$ in a way that:\\
\noindent (1) Initially set $sum=d(T_{v_1})+1$, $V^1_{T_{v_2}}=\emptyset$, $V^2_{T_{v_2}}=\emptyset$.\\
\noindent (2) Every time a new vertex $u$ is visited, if $sum+1<L^*_K$ holds, set $V^1_{T_{v_2}}=V^1_{T_{v_2}}\cup \{u\}$ and $V^2_{T_{v_2}}=\emptyset$, otherwise (if $sum+1=L^*_1+1$), set $V^1_{T_{v_2}}=V^1_{T_{v_2}} \cup{\{u\}}$ and $V^2_{T_{v_2}}=V_{T_{v_2}} \backslash V^1_{T_{v_2}}$ and stop. If $sum+1>L^*_K$ holds, set $V^1_{T_{v_2}}=\emptyset$, $V^2_{T_{v_2}}=V_{T_{v_2}}$.\\
\indent The cost to serve the clients in $T_{v_1} \cup \{v\} \cup V^1_{T_{v_2}}$ is at most
$$2C(path(r,v)) + 2C(T_{v_1}) + 2c_{vv_1} + 2C(T_{v_2}) + 2c_{vv_2}. $$
\noindent The reduced quantity from the lower bound is given by
$$2C(path(r,v)) + 2c_{vv_1} + 2C(T_{v_1}),$$
\noindent since for each edge $e \in path(r,v)$ the reduced quantity from the lower bound $LB(e)$ is $2c_e$, since the decrease of $d(T_u)$ is exactly $L_K^*$. Hence, the ratio between the cost of the tours and the reduced quantity from the lower bound is given by
$$dec_1=\frac{2C(path(r,v)) + 2C(T_{v_1}) +2c_{vv_1} + 2C(T_{v_2})+ 2c_{vv_2} }{2C(path(r,v)) +2c_{vv_1} + 2C(T_{v_1})} \le 2.$$
\noindent The clients in $T_{v_1} \cup \{v\} \cup V^1_{T_{v_2}}$ are served in the $k^{th}$ round and their demands will not be considered in the later rounds of the algorithm. Hence, we get the following lemma.
 \begin{lemma}
If $k \le K'$, the approximation ratio is at most $2$.
\end{lemma}

 \indent Now, assume $k >K'$. The algorithm first reforms the remaining tree according to the operations given above until no more operation can be applied. Then, it performs some strategies for each of the cases explained above. We will use the following simple fact:\\
 \\
$\mathbf{Fact \, 1}$. $\frac{a+c}{b+c} \le \frac{a}{b}$, for each $0<b<a$ and $c>0$.\\
 \\
\indent $\mathbf{Case \,1.}$ In this case, the algorithm focuses on arbitrary one or two p-nodes. Let $u$ be the q-node and $v,v'$ denote the two p-nodes. For the p-node $v$, we denote by $v_1, v_2, v_3$ the children of the subtree $T_{v}$, by $d_{v_1}, d_{v_2}, d_{v_3}$ their demands, and by $c_{vv_1}, c_{vv_2}, c_{vv_3}$ weights of the edges between $v$ and its children. Also, we denote by $C(path(r,u))$ the weight of the (unique) path between $r$ and $u$. The cost of the edge between $u$ and $v$ is denoted by $c_{uv}$. We have: $0 < d_{v_i}< L_K^*$, $i=1, 2, 3$; $L_K^*< d_{v_i}+d_{v_j} < 2L_K^*$, $i \neq j, i,j=1,2,3$; and so $\frac{3}{2} L_K^*< d_{v_1}+d_{v_2}+d_{v_3} < 2L_K^*$. We assume $c_{vv_1} \ge c_{vv_2} \ge c_{vv_3}$. Similarly, we denote by $v'_1, v'_2, v'_3$ the children of the subtree $T_{v'}$, by $d_{v'_1}, d_{v'_2}, d_{v'_3}$ their demands, and by $c_{v'v'_1}, c_{v'v'_2}, c_{v'v'_3}$ weights of the edges between $v'$ and its leaves. Here, we denote by $c_{uv'}$ the cost of the edge between $u$ and $v'$. In a similar way, by assumption, $0 < d_{v'_i}< L_K^*$, $i=1, 2, 3$; $L_K^* < d_{v'_i}+d_{v'_j} < 2L_K^*$, $i \neq j, i,j=1,2,3$; and $\frac{3}{2} L_K^*< d_{v'_1}+d_{v'_2}+d_{v'_3} < 2L_K^*$ hold, and $c_{v'v'_1} \ge c_{v'v'_2} \ge c_{v'v'_3}$ is assumed. \\

\indent There are two cases depending on whether $\sum_{i=1}^{i=3} d_{v_i} \ge 2L_1^*$ holds or not. If this inequality holds, the case is called subcase 1A. If it does not hold, the case is called subcase 1B.\\

\indent $Subcase \,\, 1A.$ We prepare the following strategy. \\
\indent $\mathbf{Strategy \, 2}:$ The strategy allocates two vehicles to serve the full demands of $d_{v_1}+d_{v_2}$ and a partial demand of $d_{v_3}$. The first vehicle utilizes the full demand of $d_{v_1}$ and a partial demand of $d_{v_3}$ to be filled to the maximum load of $L_1^*$. The second vehicle employs the full demand of $d_{v_2}$ and the rest demand at $v_3$ to be loaded to the utmost bar of $L_1^*$. Note that, if $d_{v_1}=L_1^*$ or $d_{v_2}=L_1^*$, the vehicles do not use the demand at $v_3$. Thus, the ratio between the cost of these vehicles and the reduced quantity from the lower bound is given by
$$dec_2=\frac{4C(path(r,u))+4c_{uv}+2c_{vv_1}+2c_{vv_2}+4c_{vv_3}}{2C(path(r,u))+2c_{uv}+2c_{vv_1}+2c_{vv_2}} \le \frac{4C(path(r,u))+4c_{uv}+8c_{vv_3}}{2C(path(r,u))+2c_{uv}+4c_{vv_3}} \le 2.$$
 \noindent The first inequality follows from fact 1 and the assumption that $c_{vv_1} \ge c_{vv_2} \ge c_{vv_3}$. We observe that the leaves $v_i$, $i=1,2,3$, are indeed the subtrees of the primitive tree $T$, and a partial demand at $v_3$ could be computed straightforwardly as we stated in strategy 1. \\

 \indent $Subcase \,\, 1B.$ Let $\sum_{i=1}^{i=3} d_{v_i} < 2L_1^*$ and $\sum_{i=1}^{i=3} d_{v'_i} < 2L_1^*$. Without loss of generality, we assume $c_{uv}+c_{vv_2} \ge c_{uv'}+c_{v'v'_2}$. We have either $\sum_{i=1}^{i=3} d_{v_i}+d_{v'_3} > 2L_1^*$, or $\sum_{i=1}^{i=3} d_{v_i}+d_{v'_2} > 2L_1^*$: since $d_{v'_3} +d_{v'_2} > L_K^*$, $\frac{3}{2} L_K^*< d_{v_1}+d_{v_2}+d_{v_3} < 2 L_K^*$, and $L_K^*\ge L_1^*$. First, assume $\sum_{i=1}^{i=3} d_{v_i}+d_{v'_2} > 2L_1^*$. We prepare the following strategy. \\
 
\indent $\mathbf{Strategy \, 3}:$ This strategy also allocates two vehicles to serve the demands of $d_{v_1}+d_{v_2}+ d_{v_3}$ and a partial demand of $d_{v'_2}$. The first vehicle utilizes the full demand of $d_{v_1}$ and a partial demand of $d_{v_3}$ to be filled to the maximum load of $L_1^*$. The second vehicle employs the full demand of $d_{v_2}$ and the rest demand at $v_3$ and a partial demand at $v'_2$ to be loaded to the utmost bar of $L_1^*$. The remaining demand of $d_{v'_2}$ will be considered in the next rounds. The ratio is as follows:
$$dec_3=\frac{4C(path(r,u))+4c_{uv}+2c_{vv_1}+2c_{vv_2}+4c_{vv_3}+2c_{uv'}+2c_{v'v'_2}}{2C(path(r,u))+4c_{uv}+2c_{vv_1}+2c_{vv_2}+2c_{vv_3}}\\
\le \frac{4C(path(r,u))+6c_{uv}+6c_{vv_2}+4c_{vv_3}}{2C(path(r,u))+4c_{uv}+4c_{vv_2}+2c_{vv_3}}\le 2.$$

\noindent The first inequality comes from the assumptions $c_{uv}+c_{vv_2} \ge c_{uv'}+c_{v'v'_2}$, $c_{vv_1} \ge c_{vv_2}$ and fact 1. \\
\indent In the case that $\sum_{i=1}^{i=3} d_{v_i}+d_{v'_3} > 2L_1^*$, the strategy is the same as strategy 3, where the role of $v'_3$ and $v'_2$ is exchanged. Since $c_{uv}+c_{vv_2} \ge c_{uv'}+c_{v'v'_2} \ge c_{uv'}+c_{v'v'_3}$, a similar way could be used to show that the ratio between the cost of the tours and the decrease in the lower bound is at most $2$. Hence, we get the following lemma.
 \begin{lemma}
For $k > K'$ and in case 1, the approximation ratio is at most $2$.
\end{lemma}

\indent $\mathbf{Case \,\, 2.}$ In this case, a q-node has no p-node. It follows from the definition of a q-node that it has at least three and at most four leaves. So, there are two cases: In subcase 2A, there exist three leaves in a q-node $u$ whose total demand is at least $2L_K^*$. In subcase 2B, there exist four leaves in a q-node whose total demand is at least $2L_K^*$. \\
\indent $Subcase \, 2A .$ In this case, a q-node $u$ has three children whose total demand is at least $2L_1^*$. This case can be considered as a special case of subcase 1A where the cost of the edge $(u,v)$ is equal to $0$. Thus, we use strategy 2.\\
\indent $Subcase \, 2B.$ In this case, a q-node $u$ has four children. This case can be considered as a special case of case 3 where the cost of the edge between $u$ and $v$ is equal to $0$. Thus, we will investigate in case 3. We conclude the following lemma.
\begin{lemma}
For $k > K'$ and in case 2A, the approximation ratio is at most $2$.
\end{lemma}

\indent $\mathbf{Case\, 3.}$ In this case, there exist (exactly) one child of p-node $v$ in a q-node $u$, and at least one child $v'$ other than $v$. Let $v_1, v_2,v_3$ be the children of the subtree $T_{v}$ with demands $d_{v_1}, d_{v_2}, d_{v_3}$, respectively. If $\sum_{i=1}^{i=3}d_{v_i} \ge 2L_1^*$, strategy 2 can be used. So, we assume $\sum_{i=1}^{i=3}d_{v_i} < 2L_1^*$. We notice here that $d(T_v)+ d_{v'}$ exceeds $2L_K^*$, since otherwise the merge operation can be applied to shift $v'$ down to the position of the child of $v$. We assume $c_{vv_1} \ge c_{vv_2} \ge c_{vv_3}$. We denote by $C(path(r,u))$ the path length between $r$ to $u$, and by $c_{uv}$ the cost of the edge between $u$ and $v$.\\

\indent  We assume $c_{uv'} > c_{uv} + c_{vv_2}$. The case where $c_{uv'} \le c_{uv} + c_{vv_2}$, can be considered as a special case of the subcase 1B in which the cost of the edge between $u$ and $v'$ is equal to $0$. Thus, we use the strategy 3, except that we exchange the role of $v'_2$ and $v'$. Since $c_{uv'} \le c_{uv}+c_{vv_2}$, a similar way could be used to show that the ratio between the cost of the tours and the reduced quantity from the lower bound is at most $2$. \\
\indent If $c_{uv'} > c_{uv} + c_{vv_2}$, we prepare the following two strategies depending on whether 
$$d_{v_1}+d_{v_2}+d_{v'} \le 2L_1^*$$
\noindent  holds or not. If the above inequality holds, the case is called subcase $3A$. If it does not hold, the case is called the subcase $3B$. \\

\indent $Subcase \,3A.$ We prepare the following strategy.\\
\indent $\mathbf{Strategy \,4}:$ The strategy employs two vehicles to serve the full demands at $v'$, $v_1$, $v_2$ and a partial demand of $d_{v_3}$. The first vehicle utilizes the demand of $d_{v'}$ and a partial demand of $d_{v_2}$ to be filled to the maximum load of $L_1^*$. The second vehicle employs the demand of $d_{v_1}$ and the remaining demand at $v_2$ and possibly a partial demand at $v_3$ to be loaded to the utmost bar of $L_1^*$. The ratio is as follows: 
$$dec_4=\frac{4C(path(r,u))+4c_{uv}+2c_{vv_1}+4c_{vv_2}+2c_{vv_3}+2c_{uv'}}{2C(path(r,u))+2c_{uv}+2c_{vv_1}+2c_{vv_2}+2c_{uv'}} \le \frac{4C(path(r,u))+4c_{uv}+10c_{vv_2}}{2C(path(r,u))+2c_{uv}+6c_{vv_2}} \le 2.$$
\noindent The first inequality comes from fact 1 and the assumptions $c_{uv'} > c_{vv_2}$, $c_{vv_1} > c_{vv_2} > c_{vv_3}$.\\

\indent $Subcase \,3B.$ If $d_{v_1}+d_{v_2}+d_{v'} > 2L_1^*$, the algorithm applies the following strategy. \\
\indent $\mathbf{Strategy \, 5}:$ The strategy employs two vehicles to serve the full demands at $v'$, $v_1$ and possibly a partial demand of $d_{v_2}$. A vehicle uses the demand of $d_{v'}$ and possibly a partial demand of $d_{v_2}$ to be filled to the maximum load of $L_1^*$. The other vehicle employs the full demand at $v_1$ and possibly a partial demand at $v_2$ to be loaded to the utmost bar of $L_1^*$. There are two cases: (a) $d_{v'}=L_1^*$, and (b) $d_{v'}<L_1^*$. In the case (a), the ratio is as follows:
$$dec_5=\frac{4C(path(r,u))+2c_{uv}+2c_{vv_1}+2c_{vv_2}+2c_{uv'}}{2C(path(r,u))+2c_{vv_1}+2c_{uv'}} \le \frac{4C(path(r,u))+4c_{vv_1}+4c_{uv'}}{2C(path(r,u))+2c_{vv_1}+2c_{uv'}} \le 2.$$
\noindent The first inequality is due to the facts $c_{uv} < c_{uv'}$, and $c_{vv_2} < c_{vv_1}$. In the case (b), the ratio is as follows:
$$dec_5 =\frac{4C(path(r,u))+4c_{uv}+2c_{vv_1}+4c_{vv_2}+2c_{uv'}}{2C(path(r,u))+2c_{uv}+2c_{vv_1}+2c_{uv'}} \le \frac{4C(path(r,u))+4c_{uv}+8c_{vv_2}}{2C(path(r,u))+2c_{uv}+4c_{vv_2}} \le 2.$$
\noindent The first inequality follows from fact 1 and the assumptions $c_{uv'} > c_{vv_2} $, $ c_{vv_1}> c_{vv_2} $. Thus, we conclude the following lemma.
\begin{lemma}
For $k > K'$ and in case 3, the approximation ratio is at most $2$.
\end{lemma}

\indent $\mathbf{Final \, round}.$ In this round, there are some leaves whose loads are equal to $L_K^*$. Also, there are three possible cases: (a) there is exactly one leaf whose load is smaller than $L_K^*$, (b) there are exactly two leaves whose loads are smaller than $L_K^*$, (c) there is a p-node $u$ for which $d(T_u) < 2L_1^*$.

\indent In the cases (a) or (b), if there is a leaf $v_j$ for which $d_{v_j}=L_1^*$, we allocate a vehicle to optimally serve the full demand at $v_j$. In the case (b), let $v_i, v_j$ be the leaves for which $d_{v_i} <L_1^*$, $d_{v_j} <L_1^*$, $d_{v_i} +d_{v_j} \ge L_1^*$. Let $c_{rv_i} \le c_{rv_j}$, we allocate a vehicle to serve the full demand at $v_j$ and a partial demand at $v_i$. Then, the ratio is  
$\frac{2c_{rv_i}+2c_{rv_j}}{2c_{rv_i}} \le 2.$

\indent Thus, we can assume there is at most one leaf whose load is smaller than $L_1^*$. In the case (c), if $d(T_u) \ge 2L_1^*$, we use strategy $2$, so we assume $d(T_u) < 2L_1^*$. 

\indent First, assume there is one p-node $u$ for which $d(T_u) <2L_1^*$, and some other leaves $v_1, v_2,...,v_h$ with demands $d_{v_i}=L_K^*$, $1 \le i \le h$. Let $u_1, u_2$, and $u_3$ be three leaves of $u$. We denote their demands by $d_{u_1}, d_{u_2}$, and $d_{u_3}$. Let $(u,u_1), (u,u_2), (u,u_3)$ be the edges connecting $u_i$, $i=1,2,3$ to their parent $u$. We denote their costs by $c_{uu_1}, c_{uu_2}, c_{uu_3}$ and denote by $c_{ru}$ the cost of the edge between $r$ and $u$. Without loss of generality, we assume $c_{uu_1} \ge c_{uu_2} \ge c_{uu_3}$. The algorithm applies the following strategy.\\

\indent $\mathbf{Strategy \, 6}:$ The strategy employs $m$ vehicles to serve the remaining demands. Each vehicle load must be equal to $L_1^*$. The $i^{th}$ vehicle, $1 \le i \le h$, serves a partial demand of $d_{v_i}$. The ${h+1}^{th}$ vehicle serves the full demand of $d_{u_1}$ and a partial demand of $d_{u_3}$, and the ${h+2}^{th}$ vehicle serves the full demand of $d_{u_2}$ and the remaining demand $d^r_{u_3}$ at $u_3$ and the remaining demands at $v_1, ... , v_l$, $l=L_1^*-(d_{u_2}+d^r_{u_3})$ so that its load is equal to $L_1^*$. The $h+i^{th}$ vehicle, $3 \le i \le m-h$, serves the remaining demands at $v_{l+1+(i-3)L_1^*} ,..., v_{l+(i-2)L_1^*}$. Note that, $v_{l+(m-h-2)L_1^*}=v_{h}$, since $\sum_{i=1}^{i=h} d_{v_i} +d(T_u)=m L_1^*$. We see the edges weights $c_{rv_i}, i=1,..,h$, $c_{ru}$, and $c_{uu_i}, i=1,2,3$, appear at most 4 times while the lower bound of each edge is exactly $2c_{rv_i}$. Thus, the ratio is as follows:
$$ dec_{6} = \frac{4 \sum_{i=1}^{h} {c_{rv_i}} + 2c_{uu_1}+ 2c_{uu_2} +4c_{uu_3}+4c_{ru}}{2\sum_{i=1}^{h} c_{rv_i}+2c_{uu_1}+ 2c_{uu_2} +2c_{uu_3}+2c_{ru}} \le 2.$$
\indent Now, assume there is a leaf with demand smaller than $L_1^*$ except $h$ leaves whose demands are equal to $L_K^*$. The strategy is the same as strategy 6, except that the ${h+1}^{th}$ vehicle serves the full demand of $d_{v_{h+1}}$ and the remaining demands at $v_1, ... , v_l$, so that its load is equal to $L_1^*$. The $i^{th}$ vehicle $h+2 \le i \le m$ serves the remaining demands in the same way as we said in the strategy 6. We see each weight $c_{rv_i}, i=1,..,h+1$, appears at most 4 times while the lower bound on each edge is $2c_{rv_i}$. Thus, the ratio is at most 2. So, we get the following lemma.
\begin{lemma}
For $k > K'$ and in the final round, the approximation ratio is at most $2$. 
\end{lemma}
\indent In the following, we study an extension to unequal demand problems.
\section{LBCVRP}
 \noindent As we stated before, the LBCVRP can be encoded in the un-splittable assignment problem. In this case, finding the fairest allocation is NP-complete \cite{Lenstra}. So, we focus on obtaining approximate solutions. We use the following natural definition of the notion of approximation. Let $L=({L_1},{L_2},...,{L_K})$ be a load allocation vector, that is sorted in non-decreasing order, and $R_L = (\frac{L_K}{L_1},...,\frac{L_2}{L_1},1)$. Let $L^*=(L^*_1, L^*_2,...,L^*_K)$ be the allocated loads in an optimal solution, sorted in non-decreasing order and $R_{L^*} = (\frac{L^*_K}{L^*_1},...,\frac{L^*_2}{L^*_1},1)$. We say $R_L$ is a $c$-approximation to $R_{L^*}$ (written $R_L \le c. R_{L^*}$), iff for each $j$ the value of the $j$th largest entry in $R_L$ is at most $c$ times the value of the $j$th largest entry in $R_{L^*}$. We say the balance/fairness of $L$ is a $c-$approximation of the balance/fairness of $L^*$, iff $R_L$ is a $c$-approximation to $R_{L^*}$. 
 \begin{definition} Let $S$ be a solution for the LBCVRP with the allocated loads $L_S$, and let $S^*$ be an optimal solution with the allocated loads $L^*$. We say the balance/fairness of $S$ is a $c-$approximation to the balance/fairness of $S^*$, iff $L_S$ is a $c$-approximation to $L^*$. 
\end{definition} 

 \indent In the following, we show there is an algorithm that finds a solution for the LBCVRP whose fairness is a $4-$approximation to the optimal.
  \subsection{LBCVRP in general metrics}
  \indent Let $\Gamma$ be a $\gamma$-approximation algorithm for the CVRP that assigns a new vehicle tour to the clients whose demands are at least $Q/2$. We show there is a polynomial-time $(\gamma +1,4)$ bi-criteria approximation algorithm for the LBCVRP that produces a solution. The total traveling cost of the produced solution is guaranteed to be about $\gamma+1$ times the optimal, and its balance is a 4-approximation to the balance of the optimal solution. The best-known result for $\gamma$ is $3.5-\frac{3}{Q}$ by the algorithm of Altinkemer and Gavish \cite{Altinkemer1987} (see lemma 9). 
 \begin{lemma}  \cite{Altinkemer1987}
There is a tour partitioning heuristic for the CVRP that provides a solution whose cost is within $3.5-\frac{3}{Q}$ times the optimal.
\end{lemma} 

\indent The Altinkemer and Gavish algorithm in \cite{Altinkemer1987} assigns a new vehicle tour for the clients whose demands are at least $Q/2$. Let $S$ be the solution found using this algorithm, and $\alpha_S$ be its balanced ratio. We have $\alpha_S \le (\frac{Q}{d_{\min }})-1$ where ${d_{\min }} = \min \{ {d_v}:v \in V \backslash \{ r\} \} $. In the following, we attempt to modify the tours in $S$, to obtain a better-balanced ratio. Define a heavy (light) tour to be a tour whose load is at least (smaller than) $Q/2$. We modify the tours by removing some nodes from a heavy tour, and appending them to the tour with the smallest load, in the same order they are appeared (see Fig. 3).  \\
\begin{figure}[ht]
\begin{center} 
\includegraphics [scale=0.28]{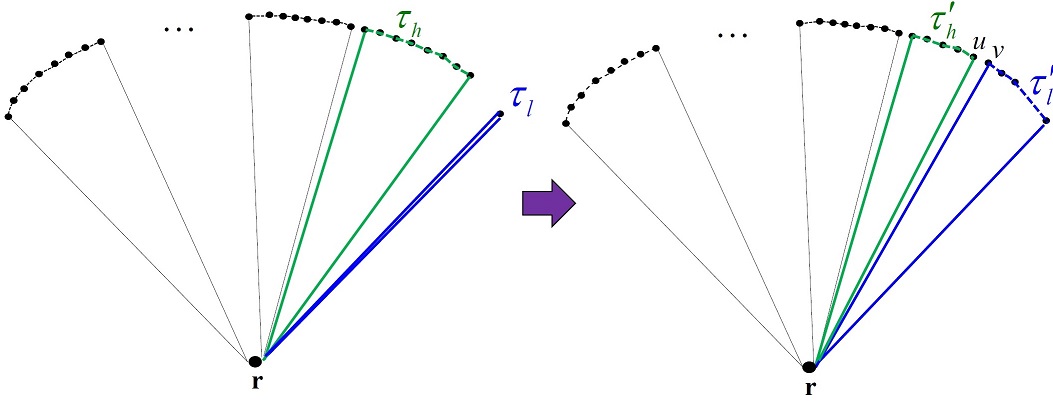} 
\end{center}
\caption{Modification of the tours.}
\label{fig3}
\end{figure}\\
\indent Let $\tau_l \in S$ be the light tour with the smallest load and $\tau_h \in S$ be a heavy tour that serves at least two clients. Modifying the tours $\tau_l$, $\tau_h$, produces two new tours  $\tau'_l$, $\tau'_h$. The edges $(r,u), (r,v)$ are added to $\tau'_h$, $\tau'_l$ and the edge $(u,v)\in \tau_h$ is removed (see Fig. 3). We see $c_{ru} + c_{rv} - c_{uv} \le 2 c_{ru} \le C_{LBCVRP}$ where $C_{LBCVRP}$ is the optimal value for the LBCVRP. Thus, the load balance would be improved at the loss of a constant factor of one in the traveling cost approximation ratio. \\
\indent Assume $L_m=(L_1,L_2,...,L_K)$ is the allocated loads in the modified solution (sorted in non-decreasing order), and $R_m=(\frac{L_K}{L_1},\frac{L_{K-1}}{L_1},...,1)$. We compare $R_m$ with a vector of ones $R=(1,1,...,1)$, since in the best case, all values of the allocated loads are equal. The smallest load is at least $Q/4$ in the modified solution (see theorem 4), so $R_m$ is a $4-$approximation of $R$. Thus, the balance of the modified solution is a $4-$approximation of the balance of the optimal solution. Moreover, we see the balanced ratio, $\alpha_m$, of the modified solution is at most $3$: $\alpha_m \le 3$.
\begin{theorem}
The smallest load is guaranteed to be at least $Q/4$, after modifying the solutions. 
\end{theorem}
\begin{proof}
\noindent We start by showing that for each tour $\tau$, $d(\tau ) \ge Q/4$. Let  $\tau_l = (r,{v_r},{v_{r + 1}},..,{v_h},r) \in S$ be the light tour and $\tau_h : = (r,{v_l},{v_{l + 1}},...,{v_k},r) \in S$ be a heavy tour that contains at least two nodes. If $d(\tau_l) = \sum\nolimits_{v \in \tau_l} {{d_v}}  \ge Q/4$, the proof is complete. So, let $d(\tau_l) < Q/4$. It is easy to see that $d(\tau_l) + d(\tau_h) > Q$ since otherwise they will be merged. Hence,
$$d(\tau_h) > \frac{3Q}{4}.$$
\noindent Let $j$ be the first index for which $d(\tau_l) + \sum\nolimits_{i= k - j}^k d_{v_i}\ge Q/4$, and let $({v_{k - j}},...,{v_{k - 1}},{v_k})$  be the vertices that are removed from $\tau_h$ joined to $\tau_l$. Let $\tau'_h$ and $\tau'_l$ be the respective new tours. We have $d(\tau'_l) \ge Q/4$ and it remains to show that $d(\tau '_h) \ge Q/4$. \\
\noindent Otherwise, let $d(\tau'_h) < Q/4$. According to the above, the index $k-j$ is the first index for which $\sum\limits_{i = k - j}^k {d_{v_i}}+ d(\tau_l) \ge Q/4$ (indeed,  $\sum\limits_{i = k - j + 1}^k d_{v_i}+ d(\tau_l) < Q/4$). Furthermore, $d_{v_{k - j}}< Q/2$ since the Altinkemer and Gavish algorithm assigns a new vehicle tour for the nodes with demand at least $Q/2$. We remind that  $d(\tau_h) + d(\tau_l) = d(\tau '_h) + d_{v_{k - j}} + \sum\limits_{i = k - j + 1}^k d_{v_i}  + d(\tau_l) > Q$. Thus,
$d_{v_{k - j}} > Q - (\sum\limits_{i = k - j + 1}^k {{d_{{v_i}}}}  + d(\tau_l) + d(\tau '_h)) > Q - Q/2= Q/2,$
which is a contradiction.\\
$\qed$
\end{proof}
\subsection{LBCVRP on Trees}
\noindent  Chandran and Raghavan in \cite{Chandran} presented a $2-$approximation algorithm for the CVRP on trees. Let $S_T=\{T_1, T_2,...,T_K\}$ be the set of subtrees found using this algorithm, and $\alpha_{S_T}$ be its balanced ratio. We have $\alpha_{S_T} \le (\frac{Q}{d_{\min }})-1$. In this section, we improve this algorithm to find a better-balanced solution for the TLBCVRP (see algorithm 4). Let $T$ be a binary tree rooted at $r$. We define a non-grandparent node to be a node that has only leaf nodes as children. For a node $v \in T$, the vertices $v_1, v_2$ are its children, and $T_v$ is the subtree hanging off $v$.

\begin{algorithm}
\caption {} 
$\mathbf{input:}$ An instance of the TLBCVRP.\\
$\mathbf{output:}$ A set of feasible vehicle tours. 
\begin{itemize}
\item[] $\mathbf{Start}$
\item[] Remove the nodes with demand at least $Q/2$ and replace a dummy vertex with zero demand.
\item[] Assign a new vehicle tour for each of the removed vertices. 
\item[] Apply the algorithm of Chandran and Raghavan \cite{Chandran}:
\begin{itemize}
\item[] $\mathbf{While}$ there is a non-grandparent node $\mathbf{do}$
\begin{itemize}
\item[] Choose a non-grandparent node $v$ and apply a bin-packing heuristic to pack the demands in $T_v=T_{v_1}\cup T_{v_2} \cup v$ into a minimal set of bins. The sum of the demands in each bin does not violate the capacity $Q$. 
\item[] Remove the nodes in the subtree $T_v$ from the graph and replace some nodes that represent the packed bins. For each bin, add one node to the tree as the child of $u$ (parent of node $v$).  
\item[] Continue this procedure until all items get packed. 
\end{itemize}
\item[] $\mathbf{end\,\,while}$
\end{itemize}
\item[] Remove some of the clients from the subtree with the highest load and append them to the tree with the smallest load:
\begin{itemize}
\item[] $T_h:=$ a heavy tree that contains at least two nodes.
\item[] $V_{T_h}:=\{u: u\in T_h\}$ (vertice of $T_h$).
\item[] $T_l:=$ the tree with the smallest load. 
\item[] Split the vertices $V_{T_h}$ into $V^1_{T_h}$ and $V^2_{T_h}$ so that $d(T_l)+d(V^1_{T_h})\ge Q/4, d(V^2_{T_h})\ge Q/4,$
\noindent where $d(V^i_{T_h})=\sum_{v \in V^i_{T_h}} d_v $, $i=1,2$. The existence of such $V^i_{T_h}$  is guaranteed by Theorem 4. 
\end{itemize}
\item[] Return the set of tours found.
\item[] $\mathbf{End}$
\end{itemize}
\end{algorithm}

 \begin{theorem} 
Algorithm 4 is a $(3,4)$ bi-criteria approximation algorithm for the TLBCVRP.
\end{theorem} 
\begin{proof} Assume that $S=\{T_1, T_2,...,T_K\}$ is the set of obtained tours before load balancing. We see that loads of the found tours are at least $Q/2$ except at most once. Thus, a proof similar to that of Chandran and Raghavan can be used to show that the total distance traveled by all the vehicles is at most twice that of the optimal. Indeed, for an edge $e=(u,v) \in T$, at least $\left\lceil \frac{\sum\nolimits{d(T_v)}}{Q} \right\rceil$ number of vehicles is required for any solution to travel $e$ since each vehicle load does not exceed $Q$. Becides, loads of the obtained tours are at least $Q/2$, so the number of vehicles that travel $e$ is at most $2\left\lceil \frac{\sum\nolimits{d(T_v)}}{Q} \right\rceil$. Thus, the total traveling cost is at most twice the lower bound, so it is at most twice the optimal.\\
\indent Modifying the loads increases the total traveling cost at most $2C(T_h) \le 2C(T) \le C_{TLBCVRP}$, where $C_{TLBCVRP}$ is the optimal value of the TLBCVRP. Hence, the load balance improves at the loss of a constant factor of one in the traveling cost approximation ratio. We conclude that there is $(3,4)$ bi-criteria approximation algorithm for the TLBCVRP that provides a solution, whose total traveling cost is at most $3$ times the optimal, and its load balance is at most $4$ times the best one.\\
$\qed$
\end{proof}
\subsection{LBCVRP as a multi-objective problem}
\noindent One of the main drawbacks using the first approach (equations (2)-(8)), is the difficulty of the existence of the feasible solutions (see section 2). To round around this difficulty, we assume a second approach. We redefine and convert the problem into a multi-objective problem with the objectives of $P_1$ and $P_2$. We combine the objectives into a single objective using a convex combination of them and consider the objective function of $P = \lambda {P_1} + (1 - \lambda ){P_2}$ for $0 \le \lambda  \le 1$. \\
\indent Our purpose in this section is to show there is an algorithm that provides a solution whose cost is within $4$ times the optimal (see theorem 6).
\begin{theorem}
Using the multi-objective approach to address the LBCVRP, there is an algorithm that provides a solution whose cost is within $4$ times the optimal.
\end{theorem}
\begin{proof}
\indent We use the results given in lemma 9. Let $S$ be the solution found using the Altinkemer and Gavish algorithm. When $P_i$ is the objective function, we use $OPT(P_i)$ to denote the optimal value. Also, $OPT$ is used to denote the optimal value of the function $P$. It could be observed that $\lambda OPT({P_1}) + (1 - \lambda )OPT({P_2}) \le OPT$. According to the lemma 9, we have $P_1(S)=C(S) \le 3.5 OPT({P_1})$. Here, $C(S)$ is the traveling cost of $S$.\\ 
\indent  For an allocation $L = ({L_1},...,{L_K})$ of loads, ${P_2}$ is defined as ${P_2} = \sum\nolimits_{i = 1}^K {L_i^2} $. Because $\sum\nolimits_{i = 1}^K {L_i}  = \sum\nolimits_{i = 1}^n {d_{v_i}}$ has a constant value, ${P_2}$ is a minimum, if ${L_i} = {L_j} = {L^*}$ for each $i \ne j$, so we have  $OPT(P_2) \ge K{L^*}^2$. In addition, ${L^*} \ge \frac{Q}{2}$, since $K{L^*} = \sum\nolimits_{i = 1}^K {{L_i}}  = \sum\nolimits_{i = 1}^n d_{v_i} $ and $K \le \frac{2\sum\nolimits_{i = 1}^n d_{v_i}}{Q}$. Note that, $\frac{2\sum\nolimits_{i = 1}^n d_{v_i}}{Q}$ is an upper bound for the required number of vehicles. Hence, $OPT({P_2}) \ge \frac{K{Q^2}}{4}$. Moreover, we have  ${P_2}(S) \le K{Q^2}$, so ${P_2}(S) \le 4OPT({P_2})$. Thus, using the function of ${P_2}$, the approximation factor of the solution found is $4$: 
$$\begin{array}{l}
\lambda {P_1}(S) + (1 - \lambda ) P_2(S) \le 3.5\lambda OPT({P_1}) + 4(1 - \lambda )OPT(P_2)\\
\hspace{3.6cm} \le 4(\lambda OPT({P_1}) + (1 - \lambda )OPT(P_2))\\
\hspace{3.6cm} \le 4OPT. 
\end{array}$$
$ \qed$
\end{proof}
 \section{Conclusion}
\noindent We studied the load balanced capacitated vehicle routing problem (LBCVRP)  to design two approximation algorithms. We presented a mathematical formulation for this problem and provided an algorithm to find the balanced loads in the equal demand version (LBECVRP). Given the balanced loads, we presented a $(1-\frac{1}{Q})\rho+\frac{3}{2}$ approximation algorithm to find a solution whose loads are balanced. Here $\rho$ is the factor of approximation for the TSP. When the graph is a tree, the total traveling cost is at most $2.5-\frac{1}{Q}$ times the optimal, since an optimal solution can be found (in polynomial time) for the TSP on a tree. We showed there is an improved $2-$approximation algorithm in this case. \\
\indent The direction of future research is to study the LBECVRP on Euclidean metrics. Theoretically, we are interested to know whether the LBECVRP can be solved more efficiently on Euclidean spaces. \\
\indent In general problem of the LBCVRP, we focused on obtaining approximate solutions since deciding whether there is a feasible solution is NP-complete. We presented a $(4.5-\frac{3}{Q},4)$ bi-criteria approximation algorithm. The approximation factor for the traveling cost of the obtained solution is $4.5-\frac{3}{Q}$, and the balance of the obtained loads is 4-approximation of the optimal balance. When the metric space is a tree, we showed there is a $(3,4)$ bi-criteria approximation algorithm for this problem.\\
\indent Furthermore, we investigated the LBCVRP from another viewpoint as a multi-objective problem. We showed there is an algorithm that provides a $4-$approximation.
\begin{acknowledgements}
The authors thank Amirkabir University of Technology for facilities.
\end{acknowledgements}
$\mathbf{Conflict\,\,of\,\, interest:}$ The authors declare that they have no conflict of interest.



\end{document}